\documentclass{ws-procs9x6}

\renewcommand{\i}{\ri}

\newcommand{\fm}{{\frak M}}

\newcommand{\ra}{\rightarrow}

\newcommand{\bra}{\langle} 
\newcommand{\ket}{\rangle}

\newcommand{\be}{\begin{equation}}
\newcommand{\ee}{\end{equation}}
\newcommand{\bea}{\begin{eqnarray}}
\newcommand{\eea}{\end{eqnarray}}

\newcommand{\ce}{{\cal E}}

\newcommand{\cc}{{\cal C}}

\newcommand{\cs}{{\cal S}}

\newcommand{\mqn}{M_{Q_n}}
\newcommand{\e}{{\rm e}}
\newcommand{\p}{{\rm p}}
\renewcommand{\d}{{\rm d}}

\newcommand{\grintl}{[\kern-.18em [}
\newcommand{\grintr}{]\kern-.18em ]}
\newcommand{\ds}{\displaystyle}

\newcounter{resultcounter}[section]

\newtheorem{thm}[resultcounter]{Theorem}
\newtheorem{lem}[resultcounter]{Lemma}
\newtheorem{cor}[resultcounter]{Corollary}

\def\bed{\begin{definition}}
\def\eed{\end{definition}}

\def\one{{\mathchoice {\rm 1\mskip-4mu l} {\rm 1\mskip-4mu l} {\rm 1\mskip-4.5mu l} {\rm 1\mskip-5mu l}}}

\def\proof{\noindent{\bf Proof.}\ \ }

 \def\cB{{\cal B}} \def\cC{{\cal C}}
 \def\cE{{\cal E}} \def\cF{{\cal F}}
 \def\cH{{\cal H}}

\def\cS{{\cal S}}

\newcommand{\N}{{\mathbb N}}

\newcommand{\C}{{\mathbb C}}

\newcommand{\E}{{\mathbb E}}
\renewcommand{\P}{{\mathbb P}}

\newcommand{\cme}{{\cal M}_{(E)}}

\def\proof{\noindent{\bf Proof.}\ \ }
\def\qed{\hfill $\Box$\medskip}

\newcommand{\ri}{{\rm i}}
\newcommand{\fer}[1]{(\ref{#1})}
\newcommand{\scalprod}[2]{\left\langle {#1}, {#2}\right\rangle}
\newcommand{\bbbone}{\mathchoice {\rm 1\mskip-4mu l} {\rm 1\mskip-4mu l}
{\rm 1\mskip-4.5mu l} {\rm 1\mskip-5mu l}}
\newcommand{\po}{\overline\omega}

\begin{document}
\title{Repeated Interaction Quantum Systems: Deterministic and Random}
\author{Alain Joye}
\address{Institut Fourier\\
Universit\'e de Grenoble\\ BP 74\\
38402 Saint Martin d'H\`eres, France}

\begin{abstract}
This paper gives an overview of recent results concerning the long time dynamics of repeated interaction quantum systems in a deterministic and random framework. We describe 
the non equilibrium steady states (NESS) such systems display and we present, as a macroscopic consequence, a second law of thermodynamics these NESS give rise to. We also explain in some details the analysis of products of certain random matrices underlying  this dynamical problem.
\end{abstract}


\bodymatter

\section{Introduction and  Model}

A repeated interaction quantum system consists of a reference quantum subsystem $\cS$ which interacts successively with the elements $\cE_m$ of a chain  $\cC=\cE_1+\cE_2+\cdots$ of independent quantum systems. At each moment in time, $\cS$ interacts precisely with one $\cE_m$ ($m$ increases as time does), while the other elements in the chain evolve freely according to their intrinsic dynamics. The complete evolution is described by the intrinsic dynamics of $\cS$ and of all the $\cE_m$, plus an interaction between $\cS$ and $\cE_m$, for each $m$. The latter is characterized by an interaction time $\tau_m>0$, and an interaction operator $V_m$ (acting on $\cS$ and $\cE_m$); during the time interval $[\tau_1+\cdots+\tau_{m-1}, \tau_1+\cdots+\tau_{m})$, $\cS$ is coupled to $\cE_m$ only via $V_m$. 
Systems with this structure are important from a physical point of view, since they arise naturally as models for fundamental experiments on the interaction of matter with quantized radiation. As an example, the ``One atom maser'' provides an experimental setup in which the system $\cS$ represents a mode of the electromagnetic field, whereas the elements $\cE_k$ describe atoms injected in the cavity, one by one, which interact with the field during their flight in the cavity. After they leave the cavity, the atoms encode some properties of the field which can be measured on these atoms \cite{MWM} , \cite{WVHW}. For  repeated interaction systems  considered as {\it ideal}, i.e. such that all atoms are identical with identical interactions and times of flight through the cavity, corresponding mathematical analyses are provided in \cite{WBKM} , \cite{bjm}. To take into account the unavoidable fluctuations in the experiment setup used to study these repeated interaction systems, modelizations incorporating randomness have been proposed and studied in \cite{bjm2} and \cite{bjm3}. With a different perspective, repeated quantum interaction models also appear naturally in the mathematical study of modelization of open quantum systems by means of quantum noises, see \cite{AP} and references therein. 
Any (continuous) master equation governing the dynamics of states on a 
system $\cS$ can be viewed as the projection of a unitary evolution driving the system $\cS$ and a field of quantum noises in interaction.  
It is shown in \cite{AP} how to recover 
such continuous models as some delicate limit of a discretization given by a repeated quantum interaction model. Let us finally mention  \cite{pillet85} for results of a similar flavour in a somewhat different framework.

\medskip

Our goal is to  present the results of the papers  \cite{bjm} , \cite{bjm2} and \cite{bjm3} on (random) repeated interaction quantum systems, which focus on the long time behaviour of these systems.

\medskip

Let us describe the mathematical framework used to describe these quantum dynamical systems.
According to the fundamental principles of quantum mechanics, states
of the systems $\cS$ and $\cE_m$ are given by normalized vectors (or
density matrices) on Hilbert spaces $\cH_\cS$ and $\cH_{\cE_m}$,
respectively, \cite{AJP,BR}\footnote{A normalized vector $\psi$ defines a ``pure'' state $A\mapsto \scalprod{\psi}{A\psi}={\rm Tr}(\varrho_\psi A)$, where $\varrho_\psi=|\psi\rangle\langle\psi|$. A general ``mixed'' state is given by a density matrix $\varrho =\sum_{n\geq 1}p_n\varrho_{\psi_n}$, where the probabilities $p_n\geq 0$ sum up to one, and where the $\psi_n$ are normalized vectors.}. We assume that $\dim\cH_\cS<\infty$, while $\dim\cH_{\cE_m}$ may be infinite. Observables $A_\cS$ and $A_{\cE_m}$ of the systems $\cS$ and $\cE_m$ are bounded operators forming {\it
  von Neumann algebras} $\fm_\cS\subset \cB(\cH_\cS)$ and
$\fm_{\cE_m}\subset \cB(\cH_{\cE_m})$. They evolve according to the {\it Heisenberg
  dynamics} ${\mathbb R}\ni t\mapsto \alpha^t_\cS(A_\cS)$ and ${\mathbb
  R}\ni t\mapsto \alpha^t_{\cE_m}(A_{\cE_m})$, where
$\alpha^t_\cS$ and $\alpha^t_{\cE_m}$ are $*$-automorphism groups of
$\fm_\cS$ and $\fm_{\cE_m}$, respectively, see
e.g. \cite{BR}. We now introduce distinguished {\it reference states}, given by vectors $\psi_{\cS}\in\cH_\cS$ and $\psi_{\cE_m}\in \cH_{\cE_m}$. Typical choices for $\psi_\cS$, $\psi_{\cE_m}$ are equilibrium (KMS) states for the dynamics $\alpha^t_\cS$, $\alpha^t_{\cE_m}$, at inverse temperatures $\beta_\cS$, $\beta_{\cE_m}$. The Hilbert space of states of the total system is the tensor product 
\[
\cH=\cH_\cS\otimes\cH_\cC,
\]
where $\cH_\cC=\bigotimes_{m\geq 1}\cH_{\cE_m}$, and where the infinite product is taken with respect to $\psi_\cC=\bigotimes_{m\geq 1}\psi_{\cE_m}$. The non-interacting dynamics is the product of the individual dynamics, defined on the algebra $\fm_{\cS}\bigotimes_{m\geq 1}\fm_{\cE_m}$ by $\alpha_\cs^t\bigotimes_{m\geq 1}\alpha_{\cE_m}^t$. It will prove  useful to consider the dynamics in the {\it Schr\"odinger picture}, i.e. as acting on vectors in $\cH$. To do this, we first implement the dynamics via unitaries, satisfying 
\begin{equation}
\alpha_\#^t(A_\#) = \e^{\ri t L_\#} A_\# \e^{-\ri t L_\#},\ t\in{\mathbb R},\ \ \mbox{and $L_\#\psi_\#=0$},
\label{m1}
\end{equation}
for any $A_\#\in {\frak M}_\#$, where $\#$ stands for either $\cS$ or $\cE_m$. The self-adjoint operators $L_\cS$ and $L_{\cE_m}$, called {\it Liouville operators} or ``positive temperature Hamiltonians'', act on $\cH_\cS$ and $\cH_{\cE_m}$, respectively. The existence and uniqueness of $L_\#$ satisfying \fer{m1} is well known, under general assumptions on the reference states $\psi_\#$ \cite{BR}. We require these states to be {\it cyclic} and {\it separating}. In particular, \fer{m1} holds if the reference states are equilibrium states. Let $\tau_m>0$ and $V_m\in\fm_{\cS}\otimes\fm_{\cE_m}$ be the interaction time and interaction operator associated to $\cS$ and $\cE_m$. We define the (discrete) repeated interaction Schr\"odinger dynamics of a state vector $\phi\in\cH$, for $m\geq 0$, by 
\begin{equation}
U(m)\phi = \e^{-\ri \widetilde L_m}\cdots\e^{-\ri \widetilde L_2}\e^{-\ri \widetilde L_1}\phi,
\label{m2}
\end{equation}
where 
\begin{equation}
\widetilde L_k = \tau_k L_k+ \tau_k\sum_{n\neq k} L_{\cE_n}
\label{m3}
\end{equation}
describes the dynamics of the system during the time interval $[\tau_1+\cdots+\tau_{k-1},\tau_1+\cdots+\tau_k)$, which corresponds to the time-step $k$ of our discrete process. Hence $L_k$ is
\begin{equation}
L_k=L_\cS +L_{\cE_k} + V_k,
\label{m4}
\end{equation}
acting on $\cH_\cS\otimes\cH_{\cE_k}$. We understand that the operator $L_{\cE_n}$ in \fer{m3} acts nontrivially only on the $n$-th factor of the Hilbert space $\cH_\cC$ of the chain. As a general rule, we will ignore tensor products with the identity operator in the notation.

A state $\varrho(\cdot)={\rm Tr}(\rho\, \cdot\,)$ given by density matrix $\rho$ on $\cH$ is called a {\it normal state}. 
Our goal is to understand the large-time asymptotics ($m\rightarrow \infty$) of expectations
\begin{equation}
\varrho\left(U(m)^* O U(m)\right)=\varrho(\alpha^m(O)),
\label{m5}
\end{equation}
for normal states $\varrho$ and certain classes of observables $O$ that we specify below. We denote the (random) repeated interaction dynamics by
\begin{equation}
\alpha^m(O) = U(m)^* O U(m).
\label{m5.1}
\end{equation}

\subsection{Van Hove Limit Type Results}

A first step towards understanding the dynamics of repeated interaction quantum systems  reduced to the reference system ${\cal S}$ was performed in the work \cite{aj}. This paper considers {\it Ideal Repeated Quantum Interaction Systems} which are characterized by identical elements ${\cal E}_k\equiv {\cal E}$ in the chain ${\cal C}$, constant interaction times $\tau_k\equiv \tau$  and identical interaction operators $V_k\equiv V\in  \fm_{\cS} \otimes  \fm_{\cE}  $ between  ${\cal S}$ and the elements  ${\cal E}$ of the chain.
In this setup, a Van Hove type analysis of the system is presented, in several regimes, to describe the dynamics of observables on $\cS$ in terms of a Markovian evolution equation of Lindblad type.  Informally, 
the simplest result of \cite{aj} reads as follows. Assume the interaction operator  
$V$ is replaced by $\lambda V$, where $\lambda>0$ is a coupling constant, and let $m$, the number of interactions during the 
time $T=m\tau$, scale like $m\simeq t/\lambda^2$, where $0\leq t <\infty$ and $\tau$ are fixed. Assume all elements of the chain are in a same thermal state at temperature $\beta$. 
Then, the weak coupling limit $\lambda\ra 0$ of the evolution of  any observable $O$  acting on $\cS$ obtained by tracing out the chain degrees of freedom from the evolution (\ref{m5.1}) satisfies, after removing a trivial free evolution, a continuous Lindblad type evolution equation in $t$. The temperature dependent generator is explicitely obtained from the interaction operator $V$ and the free dynamics. The asymptotic regimes of the parameters $(\lambda, \tau)$ characterized by $\tau\ra 0$ and $\tau\lambda^2\leq 1$ are also covered in \cite{aj}, giving rise to different Lindblad generators which all commute with the free Hamiltonian on $\cS$. The critical situation, where $\tau \ra 0$ with $\tau\lambda^2=1$  yields a quite general Lindblad generator, without any specific symmetry. In particular, it shows that any master equation driven by Lindblad operator, under reasonable assumptions, can
be viewed as a Van Hove type limit of a certain explicit repeated interaction quantum system.

\medskip
By contrast, the long time limit results obtained in  \cite{bjm} , \cite{bjm2} and \cite{bjm3} that we present here are obtained without  rescaling any coupling constant or parameter, as is usually the case with master equation techniques. It is possible to do without these approximations, making use of the structure of repeated interaction systems only, as we now show.

\section{Reduction to Products of Matrices}\label{sonly}

We first link the study of the dynamics to that of a product of reduced dynamics operators. In order to make the argument clearer, we only consider the expectation of an observable $A_\cS\in\fm_\cS$, and we take the initial state of the entire system to be given by the vector \begin{equation}
\psi_0=\psi_\cS\otimes\psi_\cC, 
\label{m0}
\end{equation}
where the $\psi_\cS$ and $\psi_\cC$ are the reference states introduced above. 
We'll comment on the general case below. The expectation of $A_\cS$ at the time-step $m$ is 
\begin{equation}
\scalprod{\psi_0}{\alpha^m(A_\cS)\psi_0} = \scalprod{\psi_0}{P\e^{\ri\widetilde L_1}\cdots \e^{\ri\widetilde L_m}A_\cS\, \e^{-\ri\widetilde L_m}\cdots \e^{-\ri\widetilde L_1}P\psi_0},
\label{m6}
\end{equation}
where we introduced  
\begin{equation}
P=\bbbone_{\cH_\cS}\bigotimes_{m\geq 1} P_{\psi_{\cE_m}},
\label{m7}
\end{equation}
the orthogonal projection onto $\cH_\cS\otimes {\mathbb C} \psi_\cC$. 
A first important ingredient in our analysis is the use of {\it $C$-Liouvilleans} introduced in  \cite{jp2002} ,
which are operators $K_k$ defined by the properties
\begin{eqnarray}
\e^{\ri\widetilde L_k} A \e^{-\ri\widetilde L_k} &=&\e^{\ri K_k} A \e^{-\ri K_k},\label{m8}\\
K_k\,\psi_\cS\otimes \psi_\cC&=&0, \label{m8.1}
\end{eqnarray}
where $A$ in \fer{m8} is any observable of the total system. The identity \fer{m8} means that the operators $K_k$ implement the same dynamics as the $\widetilde L_k$ whereas relation \fer{m8.1} selects a unique generator of the dynamics among all operators which satisfy \fer{m8}. The existence of operators $K_k$ satisfying \fer{m8} and \fer{m8.1} is rooted to the Tomita-Takesaki theory of von Neumann algebras, c.f. \cite{jp2002} and references therein. It turns out that the $K_k$ are non-normal operators on $\cH$, while the $\widetilde L_k$ are self-adjoint. Combining \fer{m8} with \fer{m6} we can write 
\begin{equation}
\scalprod{\psi_0}{\alpha^m(A_\cS)\psi_0} = \scalprod{\psi_0}{P\e^{\ri K_1}\cdots\e^{\ri K_m} P A_\cS\,\psi_0}.
\label{m9}
\end{equation}
A  second important ingredient of our approach is to realize that the {\it independence} of the sub-systems $\cE_m$ implies the relation
\begin{equation}
P\e^{\ri K_1}\cdots\e^{\ri K_m} P = P\e^{\ri K_1}P \cdots P\e^{\ri K_m}P.
\label{m10}
\end{equation}  Identifying $P\e^{\ri K_k}P$ with an operator $M_k$ on $\cH_\cS$, we thus obtain from \fer{m9} and \fer{m10},
\begin{equation}
\scalprod{\psi_0}{\alpha^m(A_\cS)\psi_0} = \scalprod{\psi_\cS}{M_1\cdots M_m A_\cS\, \psi_\cS}.
\label{m11}
\end{equation}
It follows from \fer{m8.1}
that $M_k\psi_\cS=\psi_\cS$, for all $k$, and,
because the operators $M_k=P\e^{\ri K_k}P$ implement a unitary dynamics, we
show (Lemma \ref{contraction}) that the $M_k$ are always contractions for some
suitable norm $|||\cdot |||$ on ${\mathbb C}^d$. This motivates the following\vspace{.0cm}\\

\noindent{\bf Definition:} Given a vector $\psi_\cS\in {\mathbb C}^d $ and a norm on 
$|||\cdot |||$ on ${\mathbb C}^d$,  we call {\it Reduced Dynamics Operator} any
matrix which is a contraction for $|||\cdot |||$ and  leaves $\psi_\cS$ invariant.\vspace{.2cm}

\noindent{\it Remark:} In case all couplings between $\cS$ and $\cE_k$ are absent, $V_k\equiv0$, 
$M_k=e^{i\tau_k L_\cS}$ is unitary and admits 1 as a degenerate eigenvalue.\\

We will come back on the properties of reduced dynamics operators (RDO's, for short) below. Let us emphasize here that the reduction process to product of RDO's is free from any approximation, so that the set of matrices $\{M_k=P\e^{\ri K_k}P\}_{k\in \mathbb N}$ encodes the complete dynamics. In particular, we show, using the cyclicity and separability of the reference vectors $\psi_\cS, \psi_{\cE_{k}}$, that the evolution of any normal state, not only $\bra \psi_0 , \ \cdot \ \psi_0\ket$, can be understood completely in terms of the product of these RDO's.

We are now in a position to state our main results concerning the asymptotic dynamics of normal states $\varrho$ acting on certain observables $O$. These result involve a spectral hypothesis which we introduce in the next \\

\noindent{\bf Definition:}
Let $\cme$ denote the set of reduced dynamics operators whose spectrum $\sigma(M)$ 
satisfies $\sigma(M)\cap \{z \in {\mathbb C} \ |\ |z|=1\}=\{1\}$ and $1$ is  simple eigenvalue.
\vspace{.2cm}


We shall denote by $P_{1,M}$ the spectral projector of a matrix $M$ corresponding to the eigenvalue 1. As usual, if the eigenvalue 1 is simple, with corresponding normalized eigenvector $\psi_\cS$, we shall write $P_{1,M}=|\psi_\cS\ket\bra\psi |$ for some 
$\psi$ s.t. $\bra\psi|\psi_\cS\ket=1$.

\section{Results}

\subsection{Ideal Repeated Interaction Quantum System }

We consider first the case of {\em Ideal Repeated Interaction Quantum Systems}, characterized by 
\bea
\cE_{k}&=&\cE, \ L_{\cE_{k}}=L_{\cE}, \  V_k=V, \ \tau_k=\tau \ \ \ \ \mbox{for all} \ \ k\geq 1,\nonumber \\ 
M_k&=& M, \  \ \forall k\geq 1.
\eea
\begin{thm}
\label{thm0}
{}\ Let $\alpha^{n}$ be the repeated interaction dynamics determined by one {\em RDO} 
$M$. Suppose that $M\in\cme$ so that  $P_{1,M}=|\psi_\cS\ket\bra\psi |$. Then, for any 
 $0<\gamma<\inf_{z\in \sigma(M)\setminus\{1\}}(1-|z|)$,
any  normal state $\varrho$, and any $A_\cs\in\fm_\cS$,
\begin{equation}
\varrho\left( \alpha^{n}(A_\cS)\right)=\scalprod{\psi}{A_\cS \psi_\cS}+
O(e^{-\gamma n}).
\label{m22}
\end{equation}
\end{thm}

\noindent{\it Remarks:} 1. The asymptotic state $\bra \psi | \cdot \psi_\cS\ket$ and
the exponential decay rate $\gamma$ are both determined by the spectral properties of 
the RDO $M$. 

2. On concrete examples, the verification of the spectral assumption on $M$ can be
done by rigorous perturbation theory, see \cite{bjm}. It is reminiscent of a Fermi Golden Rule type condition on the efficiency of the coupling $V$, see the remark  following the 
definiton of RDO's.

3. Other properties  of ideal repeated interaction quantum systems are discussed in \cite{bjm} , e.g. continuous time evolution and correlations.\\

For deterministic systems which are {\it not} ideal, the quantity $ \varrho\left( \alpha^{n}(A_\cS)\right)$ keeps fluctuating as $n$ increases, which,  in general, forbids convergence, see Proposition \ref{limpsi}. That's why we resort to ergodic limits in a random setup, as we now explain.

\subsection{Random Repeated Interaction Quantum System}\label{rsetup}

To allow a description of the effects of fluctuations on the dynamics of repeated interaction quantum systems, we consider the following setup.

Let $\omega\mapsto M(\omega)$ be a random matrix valued variable on ${\mathbb C}^d$ defined on 
a probability space $(\Omega,\cF,\p)$. We say that $M(\omega)$ is a {\it random reduced dynamics operator} (RRDO) if
\begin{itemize}
\item[(i)] There exists a norm $||| \cdot |||$ on $\C^d$ such that, for all $\omega$,
      $M(\omega)$ is a contraction on $\C^d$ for the norm $|||\cdot|||$. 
\item[(ii)] There exists a vector $\psi_\cS$, constant in $\omega$, such that $M(\omega)\psi_\cS=\psi_\cS$, for all $\omega$.
\end{itemize}

To an RRDO $\omega \mapsto M(\omega)$ on $\Omega$ is naturally associated a iid {\it random reduced dynamics process} (RRDP)
\be\label{rrr}
 \po\mapsto M(\omega_1)\cdots M(\omega_n),\qquad \po\in{\Omega}^{{\mathbb N}^*},
\ee 
where we define in a standard fashion a probability measure $\d\P$  on $\Omega^{\N^*}$ by
\[
\d\P =\Pi_{j\geq 1}\d \p_j, \ \ \ \mbox{where } 
\ \ \ \d \p_j\equiv \d \p, \ \  \forall j\in \N^*.
\]
We shall write the expectation of any random variable $f$ as ${\mathbb E}[f]$.\\

Let us denote by $\alpha_{\po}^{n}$, $\po\in\Omega^{{\mathbb N}^*}$, the process obtained from \fer{m5.1}, \fer{m11}, where the $M_j=M(\omega_j)$ in \fer{m11} are iid random matrices. We call $\alpha_{\po}^{n}$ the random repeated interaction dynamics determined by the RRDO $M(\omega)=P\e^{\ri K(\omega)}P$. It is the independence of the successive elements $\cE_k$ of the chain $\cC$ which motivates the assumption that the process (\ref{rrr}) be iid.

\begin{thm}
\label{thm4}
{}\ Let $\alpha_{\po}^{n}$ be the random repeated interaction dynamics determined by an RRDO $M(\omega)$. Suppose that $\p(M(\omega)\in\cme)>0$. Then there exists a set $\overline{\Omega}\subset \Omega^{\N^*}$, s.t.  ${\mathbb P}(\overline{\Omega})=1$, and s.t. for any $\po\in\overline{\Omega}$, any normal state $\varrho$ and any $A_\cs\in\fm_\cS$,
\begin{equation}
\lim_{N\rightarrow\infty} \frac 1N\sum_{n=1}^N \varrho\left( \alpha^{n}_{\po}(A_\cS)\right)=\scalprod{\theta}{A_\cS \psi_\cS},
\label{m222}
\end{equation}
where $\theta=P_{1,{\mathbb E}[M]}^*\psi_\cS$.
\end{thm}

\medskip
\noindent {\it Remarks:} 1. Our setup allows us to treat systems having various sources of randomness. For example, random interactions or times of interactions, as well as  random characteristics of the systems $\cE_m$ and $\cS$ such as random temperatures and dimensions of the $\cE_m$ and of $\cS$.

2. The asymptotic state $\scalprod{\theta}{\ \cdot \ \psi_\cS} $ is again determined by
the spectral data of a matrix, the expectation ${\mathbb E}[M]$ of the RRDO $M(\omega)$. Actually, our hypotheses imply that ${\mathbb E}[M]$ belongs to $\cme$, see below. 

3. The explicit computation of the asymptotic state, in this Theorem and in the previous one, is in general difficult. Nevertheless, they can be reached by rigorous perturbation theory, see the examples in \cite{bjm} , \cite{bjm2} and  \cite{bjm3}.

\subsection{Instantaneous Observables}

There are important physical observables  that describe exchange processes between $\cS$ and the chain $\cC$ and, which, therefore, are not represented by operators that act just on $\cS$. To take into account such phenomena, we consider the set of observables defined as follows.\vspace{.2cm}\\

\noindent{\bf Definition:} The
{\it instantaneous observables} of $\cS+\cC$ are of the form 
\begin{equation}
O= A_\cS\otimes_{j=-l}^rB_m^{(j)},
\label{intro0'}
\end{equation}
where $A_\cS\in\fm_\cS$ and $B_m^{(j)} \in \fm_{\cE_{m+j}}$. \\

\noindent Instantaneous observables can be viewed as a train of $l+r+1$ observables, roughly centered at $\cE_m$, which travel along the chain $\cC$ with time.\\

Following the same steps as in Section \ref{sonly}, we arrive at the 
following expression for the evolution of the state $\psi_0$ acting on an instantaneous observable $O$ at time $m$:
\begin{equation}
\scalprod{\psi_0}{\alpha^m(O)\psi_0} = \scalprod{\psi_0}{PM_1 \cdots M_{m-l-1} N_m(O)P\psi_0}.
\label{intro1'}
\end{equation}
Here again, $P$ is the orthogonal projection onto $\cH_\cS$, along $\psi_\cC$.  
The operator $N_m(O)$ acts on $\cH_\cS$ and has the expression (Proposition 2.4 in \cite{bjm3})
\bea
&&N_m(O)\psi_0 = \\\nonumber
& &\hspace{.4cm}P \e^{\ri\tau_{m-l} \widetilde L_{m-l}}\cdots
\e^{\ri \tau_m \widetilde L_m} (A_\cS\otimes_{j=-l}^{r} B^{(j)}_m)
\e^{-\ri\tau_m \widetilde L_m}\cdots \e^{-\ri \tau_{m-l}
\widetilde L_{m-l}} \psi_0.
\label{nmo'}
\eea

We want to analyze the asymptotics $m\rightarrow\infty$ of (\ref{intro1'}), allowing 
for randomness in the system. 
We make the following assumptions on the random instantaneous observable: 
\begin{itemize}
\item[(R1)] The operators $M_k$ are RRDO's, and we write the corresponding iid random matrices $M_k=M(\omega_k)$, $k=1,2, \cdots$, .
\item[(R2)] The random operator $N_m(O)$ is independent of the $M_k$ with $1\leq k\leq m-l-1$, and it has the form $N(\omega_{m-l},\ldots,\omega_{m+r})$, where $N:\Omega^{r+l+1}\rightarrow  {\cal B}({\mathbb C}^d)$ is an operator valued random variable.
\end{itemize}

The operator $M_k$ describes the effect of the random $k$-th interaction on $\cS$, 
as before. The random variable $N$ in (R2) does not depend on the time step $m$, which is a condition on the observables. It means that the nature of the quantities measured at time $m$ are the same. For instance, the $B^{(j)}_m$ in (\ref{intro0'}) can represent the energy of $\cE_{m+j}$, or the part of the interaction energy $V_{m+j}$ belonging to $\cE_{m+j}$, etc. Both assumptions are verified in a wide variety of physical systems: we may take random interaction times $\tau_k=\tau(\omega_k)$, random coupling operators $V_k=V(\omega_k)$, random energy levels of the $\cE_k$ encoded in $L_{\cE_k}=L_\cE(\omega_k)$, random temperatures $\beta_{\cE_k}=\beta_\cE(\omega_k)$ of the initial states of $\cE_k$, and so on.

The expectation value in any normal state of such instantaneous observables reaches an asymptotic value in the ergodic limit given in the next

\begin{thm}
\label{thmintro2} 
Suppose that $\p(M(\omega)\in\cme)\neq 0$. There exists a set $\widetilde{\Omega}\subset\Omega^{\N^*}$ of probability one 
s.t. for any $\po\in\widetilde{\Omega}$, for any instantaneous observable $O$, \fer{intro0'}, and for any normal initial state $\varrho$, we have 
\begin{equation}
\lim_{\mu\to\infty} \frac{1}{\mu} \sum_{m=1}^\mu \varrho \big( \alpha^m_{\po} (O)\big) =\scalprod{\theta}{{\mathbb E}[N]\psi_\cS},\ \ {\mathbb E}[N]\in\fm_\cS.
\label{intro-1'}
\end{equation}
\end{thm}

{\indent}{\it Remarks} 1. The asymptotic state in which one computes the expectation (w.r.t the randomness) of $N$ is the same as in Theorem \ref{thm4}, with 
$\theta=P_{1,{\mathbb E}[M]}^*\psi_\cS$.

2. In case the system is deterministic and ideal, the same result holds, dropping the 
expectation on the randomness and taking $\theta=\psi$, as in Theorem \ref{thm0}, see \cite{bjm}.

\subsection{Energy and Entropy Fluxes}

Let us consider some  macroscopic properties of the asymptotic state. The systems we consider may contain randomness, but we drop the variable $\po$ from the notation.

Since we deal with open systems, we cannot speak about its total energy; however, variations in total energy are often well defined. Using an argument of \cite{bjm} one gets a formal expression for the total energy which is constant during all time-intervals $[\tau_{m-1},\tau_m)$, and which undergoes a jump 
\be\label{def:energy}
j(m):=\alpha^m(V_{m+1}-V_m)
\ee
at time step $m$. 
Hence, the variation of the total energy between the instants $0$ and $m$ is then $\Delta E(m)=\sum_{k=1}^m j(k)$. The relative entropy of $\varrho$ with respect to $\varrho_0$, two normal states on $\fm$, is denoted by ${\rm Ent}(\varrho|\varrho_0)$. Our definition of relative entropy differs from that given in \cite{BR} by a sign, so that in our case, ${\rm Ent}(\varrho|\varrho_0)\geq 0$. For a thermodynamic interpretation of entropy and its relation
to energy, we assume for the next result that $\psi_\cS$
is a $(\beta_\cS,\alpha_\cS^t)$--KMS state on $\fm_\cS$, and that
the $\psi_{\cE_m}$ are $(\beta_{\cE_m}, \alpha_{\cE_m}^t)$--KMS
state on $\fm_{\cE_m}$, where $\beta_\cS$ is the inverse temperature of $\cS$, and $\beta_{\cE_m}$ are
random inverse temperatures of the $\cE_m$. 
Let $\varrho_0$ be the state on $\fm$
determined by the vector
$\psi_0=\psi_\cS\otimes\psi_\cc=\psi_\cS\bigotimes_m \psi_{\cE_m}$.
The change of relative entropy is denoted $\Delta S(m) := {\rm Ent} (\varrho\circ\alpha^m| \varrho_0)-{\rm Ent} (\varrho|
\varrho_0)$. This quantity can be expressed in terms of the Liouvillean and interaction operators by means of a formula proven in \cite{JP1}.

One checks that both the energy variation and the entropy variations can be expressed as instantaneous observables, to which we can apply the results of the previous Section. 
Defining the asymptotic energy and entropy productions by the limits, if they exist,
\bea
\lim_{m\to\infty} \varrho\left(\frac{\Delta E(m}{m}\right) =: \d E_+ \ \ \ \mbox{and } \ \ \
\lim_{m\to\infty} \frac{\Delta S(m)}{m} =: \d S_+,
\eea
we obtain
\begin{thm}[$2^{\rm nd}$ law of thermodyn\-amics] 
\label{thmintro4} 
Let $\varrho$ be a normal state on
  $\fm$. Then
\begin{eqnarray*} 
\d E_+ 
&=& 
\scalprod{\theta}{\E\big[ P( L_\cS+V-\e^{\i\tau L} (L_\cS+V)\e^{-\i\tau L})P\big]
\psi_\cS}\ \ a.s.\\
 \d S_+ &=&
\scalprod{\theta}{\E\big[ \beta_\cE \, P( L_\cS+V-\e^{\i\tau L} (L_\cS+V)\e^{-\i\tau L})P\big]
\psi_\cS}
\ \ a.s.
\end{eqnarray*}
The energy- and entropy productions  $\d E_+$ and $\d S_+$ are independent of the initial state $\varrho$. If $\beta_\cE$ is deterministic, i.e., $\po$-independent, then the system satisfies the second law of thermodynamics: $\d S_+ = \beta_\ce \d E_+$.
\end{thm}
{\it Remark:} There are explicit examples in which the entropy production can be obtained via rigorous perturbation theory and is proven to be strictly positive, a sure sign that the asymptotic state is a NESS, see \cite{bjm} .

As motivated by (\ref{m11}), the theorems presented in this Sections all rely on
the analysis of products of large numbers of (random) RDO's. The rest of this note is 
devoted to a presentation of some of the key features such products have.

\section{Basic Properties of RDO's}

Let us start with a result proven in  \cite{bjm}  as Proposition 2.1.
\begin{lemma}
\label{contraction} Under our general assumptions, the set of matrices $\{M_j\}_{j\in{\mathbb N}^*}$ defined in $(\ref{m11})$ satisfy $M_j\psi_\cS=\psi_\cS$, for all $j\in \N^*$. Moreover, to any $\phi\in \cH_\cS$ there corresponds a unique $A\in
\fm_\cS$ such that $\phi=A\psi_\cS$. $|||\phi|||:=\|A\|_{\cB(\cH_\cS)}$
defines a norm on $\cH_\cS$, and as operators on $\cH_\cS$ endowed with this norm, the $M_j$ are contractions for any $j\in\N^*$. 
\end{lemma}
Again,  the fact that $\psi_\cS$ is invariant under $M_j$ is a consequence of (\ref{m8.1}) and their being contractions  comes from the unitarity of the quantum evolution together with the finite dimension of ${\cal H}_\cS$. 

As a consequence of the equivalence of the norms $\|\cdot\|$ and $|||\cdot|||$, we get
\begin{cor}\label{ubp}
We have $1\in \sigma(M_j)\subset \{z \ | \ |z|=1\}$ and 
\[
\ds \sup\, \{ \|M_{j_n}M_{j_{n-1}}\cdots M_{j_1}\|, \ n\in\N^*, \, j_k\in{\mathbb N}^*\} =C_0<\infty
\]
\end{cor} 

Actually, if a set of operators satisfies the bound of the Corollary, it is always possible to construct a norm on ${\mathbb C}^d$ relative to which they are contractions, as proven in the next
\begin{lemma}
Let $R=\{M_j\in M_d({\mathbb C})\}_{j\in J}$, where $J$ is any set of indices 
and $C(R)\geq 1$ such that
\begin{equation}\label{estr}
\|M_{j_1} M_{j_2}\cdots M_{j_n}\|\leq C(R),\ \ \forall
\{{j_i}\}_{i=1,\cdots, n}\in J^n, \ \forall n\in{\mathbb N}.
\end{equation}
Then, there exists  a
norm $||| \cdot |||$ on ${\mathbb C}^d$, which depends on $R$, 
relative to which the elements of $R$ are
contractions. 
\end{lemma}

{\bf Proof:}
Let us define $T\subset M_d({\mathbb C})$ by 
\begin{equation}
T=\cup_{n\in {\mathbb N}}\cup_{(j_1, j_2, \cdots j_n)\in
  J^n}M_{j_1}M_{j_2}\cdots M_{j_n}.
\end{equation}
Obviously $R\subset T$, but the identity matrix ${\mathbb I}$ does not
necessarily belong to $T$. Moreover, the estimate (\ref{estr}) still
holds if the $M_{j_i}$'s belong to $T$ instead of $R$. 
For any $\varphi\in {\mathbb C}^d$ we set
\begin{equation}
|||\varphi |||=\sup_{M\in T\cup{\mathbb I}}\|M\varphi\|\geq \|\varphi\|,
\end{equation}
which defines a {\it bona fide} norm. Then, for any vector $\varphi$ and any element $N$ of $T$ we compute
\begin{equation}
||| N\varphi |||=\sup_{M\in T\cup{\mathbb I}}\|MN\varphi\|\leq 
\sup_{M \in T\cup{\mathbb I}}\|M\varphi\|=|||\varphi |||,
\end{equation}
from which the result follows. {\hfill  {\vrule height 10pt width 8pt depth 0pt}}\\

\noindent {\it Remark.}
If there exists a vector $\psi_S$ invariant under all elements of $R$, it 
is invariant under all elements of $T$ and satisfies 
$\|\psi_S\|=|||\psi_S|||=1$.

\section{Deterministic Results}
\label{sec:determinist}

In this section, we derive some algebraic formulae and some uniform
bounds for later purposes. Since there
is no probabilistic issue involved here, we shall therefore 
simply denote
$M_j=M(\omega_j)$. We are concerned with the product   
\be\label{def:psi}
\Psi_n:= M_1 \cdots M_n.
\ee


\subsection{Decomposition of the $M_j$}
\label{ssec:notation}

With $P_{1,M_j}$ the spectral projection of $M_j$ for the eigenvalue $1$ we define  
\be\label{defpsi}
\psi_j:= P^*_{1,j}\psi_\cS, \ \ \ \ P_j:=|\psi_\cS\ket\bra \psi_j|.
\ee
Note that $\bra \psi_j|\psi_\cS\ket=1$ so that $P_j$ is a projection and, moreover, $M_j^*\psi_j=\psi_j$.  We introduce the following decomposition of $M_j$
\be\label{struct}
M_j:=P_j+Q_jM_jQ_j,  \ \ \ \mbox{with} \ \ \ Q_j=\one-P_j.
\ee
We denote the part of $M_j$ in $Q_j\C^d$, by 
$M_{Q_j}:=Q_jM_jQ_j$. 
It easily follows from these definitions that
\bea
P_jP_k=P_k,  & & \phantom{x}Q_jQ_k=Q_j, \label{eq:ppqq}\\
Q_jP_k=0,\phantom{i} & & \phantom{x}P_kQ_j=P_k-P_j=Q_j-Q_k. \label{eq:pqqp}
\eea

\noindent {\it Remark.}
If $1$ is a simple eigenvalue, $P_{1,M_j}=P_j$ and (\ref{struct}) is a (partial) spectral decomposition of $M_j$.

\begin{proposition}
\label{prop:psiform}
For any $n$, 
\be\label{eq:psi}
\Psi_n=|\psi_\cS\ket\bra\theta_n|+M_{Q_1}\cdots \mqn,
\ee
where
\bea
\theta_n & = & \psi_n+M_{Q_n}^*\psi_{n-1}+\cdots +M_{Q_n}^*\cdots M_{Q_2}^* \psi_1 \label{eq:theta1}\\
 & = & M_n^*\cdots M_2^* \psi_1 
\label{eq:theta2}
\eea
and where $\bra\psi_\cS,\theta_n\ket=1$.
\end{proposition}

\proof Inserting the decomposition (\ref{struct}) into (\ref{def:psi}), and using \fer{eq:ppqq}, \fer{eq:pqqp}, we have
\be
\Psi_n= \sum_{j=1}^n P_j M_{Q_{j+1}}\cdots M_{Q_n} +M_{Q_1}\cdots \mqn.\nonumber
\ee
Since $P_j=|\psi_\cS\ket\bra \psi_j|,$ this proves (\ref{eq:psi}) and (\ref{eq:theta1}). From (\ref{eq:pqqp}), we obtain for any $j,k$,
\be
\label{mqm}
M_{Q_j}M_{Q_k}=M_{Q_j}M_k=Q_jM_jM_k.
\ee
Hence, $\Psi_n = P_1 M_1\cdots M_n +Q_1 M_1\cdots M_n = |\psi_\cS\ket\bra M_n^*\cdots M_2^* \psi_1| + M_{Q_1}\cdots \mqn$, 
which proves (\ref{eq:theta2}).
\qed


\subsection{Uniform Bounds}\label{ssec:unifbound}

The operators $M_j$, and hence the product $\Psi_n$, are contractions on $\C^d$ for the norm $|||\cdot |||$.  In order to study their asymptotic
behaviour, we need some uniform bounds on the $P_j, Q_j,\ldots$ 
Recall  that $\|\psi_\cS\|=1$.

\begin{proposition}\label{prop:unifbound} Let $C_0$ be as in Corollary \ref{ubp}. Then, the following bounds hold
\begin{enumerate}
\item  For any $n\in \N^*$,  $\|\Psi_n\|\leq C_0$.
\item For any $j\in\N^*$, $\|P_j\|=\|\psi_j\| \leq C_0$ and $\|Q_j\|\leq 1+C_0$.
\item $\ds \sup\, \{\|M_{Q_{j_n}}M_{Q_{j_{n-1}}}\cdots M_{Q_{j_1}}\|, \ n\in\N^*,\, j_k\in{\mathbb N}^*\} \leq C_0(1+C_0)$.
\item For any $n\in\N^*$, $\|\theta_n\|\leq C_0^2$.
\end{enumerate}
\end{proposition}

\proof It is based on Von Neumann's ergodic Theorem, which states that
\[
P_{1,M_j}=\lim_{N\ra\infty}\frac{1}{N}\sum_{k=0}^{N-1}M_j^k.
\]
The first two estimate easily follow, whereas the third makes use of
(\ref{mqm}) to get 
$M_{Q_{j_n}}M_{Q_{j_{n-1}}}\cdots M_{Q_{j_1}}=Q_{j_n}M_{j_n}M_{j_{n-1}}\cdots M_{j_1},$
so that
\[
\|M_{Q_{j_n}}M_{Q_{j_{n-1}}}\cdots M_{Q_{j_1}}\|\leq \|Q_{j_n}\|C_0\leq C_0(1+C_0).
\]
Finally, (\ref{eq:theta2}) and the above estimates yield $\|\theta_n\|\leq C_0\|\psi_1\|\leq C_0^2$.
\qed


\subsection{Asymptotic Behaviour}\label{ssec:asympt}

We now turn to the study of the asymptotic behaviour of $\Psi_n$, starting with the simpler case of {\em Ideal Repeated Interaction Quantum Systems}. 

That means we assume 
\be
M_k= M, \  \ \forall k\geq 1.
\ee
If $1$ is a simple eigenvalue of $M$, then $P_{1,M}=|\psi_\cS\ket\bra \psi|$,
for some $\psi$ s.t. $\bra\psi |\psi_\cS\ket=1$,
and 
\be
\Psi_n=M^n=|\psi_\cS\ket\bra \psi|+M_Q^n
\ee 
Further, if all other eigenvalues of $M$ belong to the open unit disk,  $M_Q^n$ converges exponentially fast to zero as $n\ra \infty$.

Consequently, denoting by $\mbox{spr}(N)$ the spectral radius of $N\in M_d({\mathbb C})$,
\begin{lem} If the {\em RDO} $M$ belongs to $\cme$,
\be
\Psi_n=|\psi_\cS\ket\bra \psi|+O(e^{-\gamma n}),
\ee 
for all $0<\gamma<1-\mbox{\em spr}(M_Q)$. 
\end{lem}

Two things are used above, the decay of $M_Q^n$ and the fact that $\theta_n=\psi$ is constant, see (\ref{eq:psi}). The following result shows that in general, if one knows {\em a priori} that the products of $M_{Q_j}$'s in (\ref{eq:psi}) goes to zero, $\Psi_n$ converges if and only if $P_n=|\psi_\cs\ket\bra\psi_n |$, does.

\begin{proposition}
\label{limpsi} 
Suppose that $\lim_{n\to\infty} \sup \{\|M_{Q_{j_n}}\cdots M_{Q_{j_1}}\|, \ j_k\in\N^*\}=0$. Then $\theta_n$ converges if and only if $\psi_n$ does. If they exist, these two limits coincide, and thus   
\[
\lim_{n\ra\infty}\Psi_n=|\psi_\cS\ket\bra \psi_\infty|,
\]
where $\psi_\infty=\lim_{n\rightarrow\infty} \psi_n$. Moreover, $|\psi_\cS\ket\bra \psi_\infty|$ is a projection.
\end{proposition}

In general, we cannot expect  pointwise convergence of the $\theta_n$, but we can consider an ergodic average of $\theta_n$ instead. This is natural in terms of dynamical systems, a fluctuating system does not converge.

The previous convergence results relies on the
decay of the product of operators $M_{Q_j}$. 
Conditions ensuring this are rather strong. However, Theorem \ref{thm1} below shows that in the {\it random setting}, a similar exponential decay holds under rather weaker assumptions.

\section{Random Framework}

\subsection{Product of Random Matrices}

We now turn to the random setup in the framework of Section \ref{rsetup}. For $M(\omega)$ an RRDO, with probability space $(\Omega,\cF,\p)$, 
we consider the RRDP on ${\Omega}^{{\mathbb N}^*}$ given by
\[
\Psi_n(\po):=M(\omega_1)\cdots M(\omega_n),\qquad \po\in{\Omega}^{{\mathbb N}^*}.
\] 

We show that $\Psi_n$ has a decomposition into an exponentially decaying part and a fluctuating part. 
Let $P_{1}(\omega)$ denote the spectral projection of $M(\omega)$ corresponding to the eigenvalue one ($\dim P_1(\omega)\geq 1$), and let $P^*_1(\omega)$ be its adjoint operator. Define
\begin{equation}
\psi(\omega) := P_1(\omega)^*\psi_\cS,
\label{m13}
\end{equation}
and set
\bea
P(\omega)=|\psi_\cs\rangle \langle\psi(\omega)|, \ \ \ 
Q(\omega)=\one-P(\omega).
\eea
The vector $\psi(\omega)$ is normalized as $\scalprod{\psi_\cS}{\psi(\omega)}=1$. We decompose $M(\omega)$ as
\begin{equation}
M(\omega) = P(\omega)+Q(\omega)M(\omega)Q(\omega)=: P(\omega) +M_Q(\omega).
\label{m14}
\end{equation}
Taking into account this decomposition, we obtain (c.f. Proposition \ref{prop:psiform})
\begin{equation}
\Psi_n(\po):=M(\omega_1)\cdots M(\omega_n) = |\psi_\cS\rangle\langle\theta_n(\po)| + M_Q(\omega_1)\cdots M_Q(\omega_n),
\label{m15}
\end{equation}
where $\theta_n(\po)$ is the Markov process
\bea
\theta_n(\po) &=&M^*(\omega_n)\cdots M^*(\omega_2)\psi(\omega_1)\\ \nonumber&=&\psi(\omega_n)+
M_{Q}^*(\omega_n)\psi(\omega_{n-1})+\cdots +M_{Q}^*(\omega_n)\cdots M_{Q}^*(\omega_2) \psi(\omega_1) ,
\label{m16}
\eea
$M^*(\omega_j)$ being the adjoint operator of $M(\omega_j)$. We analyze the two parts in the r.h.s. of \fer{m15} separately.

\begin{thm}[Decaying process]
\label{thm1}
Let $M(\omega)$ be a random reduced dynamics operator. Suppose that $\p(M(\omega)\in\cme)>0$.
Then there exist a set $\Omega_1\subset\Omega^{\N^*}$ and constants $C,\alpha>0$ s.t. ${\mathbb P}(\Omega_1)=1$ and s.t. for any $\po\in\Omega_1$, there exists a random variable $n_0(\omega)$ s.t. for any $n\geq n_0(\po)$,
\begin{equation}
\| M_{Q}(\omega_1)\cdots M_{Q}(\omega_n)\|\leq C\e^{-\alpha n},
\label{m17}
\end{equation}
and ${\mathbb E}[e^{\alpha n_0}]<\infty$. Moreover, ${\mathbb E}[M]\in\cme$.
\end{thm}

\noindent {\it Remarks.}  1.  The sole condition of $M$ having an arbitrarily small, non-vanishing probability to be in $\cme$ suffices to guarantee the exponential decay of the product in \fer{m17} and that ${\mathbb E}[M]$ belongs to $\cme$.

2. Actually, ${\mathbb E}[M]\in\cme$ is a consequence of $\mbox{spr}({\mathbb E}[M_Q])<1$, which comes as a by product of the proof of Theorem \ref{thm1}. 
From the identities
\be \label{idmoy}
{\mathbb E}[M]=|\psi_\cS\ket\bra{\mathbb E}[\psi]|+{\mathbb E}[M_Q], \ \ \ 
 \bra{\mathbb E}[\psi]| \psi_\cS\ket=1,  \ \ \ {\mathbb E}[M_Q]\psi_\cS=0,
\ee which {\it do not } correspond to a (partial) spectral decomposition of ${\mathbb E}[M]$, and this estimate, we get
\bea
{\mathbb E}[M]^n&=&|\psi_\cS\ket\bra\ {\mathbb E}[\psi]+{\mathbb E}[M_Q]^*{\mathbb E}[\psi]+\cdots {{\mathbb E}[M_Q]^*}^{n-1}{\mathbb E}[\psi]\ |+{{\mathbb E}[M_Q]^*}^{n}\nonumber\\ \label{pem}
&{\rightarrow \atop n\ra\infty}& |\psi_\cS\ket\bra({\mathbb I}-{\mathbb E}[M_Q]^*)^{-1}{\mathbb E}[\psi]|\equiv P_{1, {\mathbb E}[M]}.
\eea

3. Our choice \fer{m13} makes $\psi(\omega)$ an eigenvector of $M^*(\omega)$. Other choices of (measurable) $\psi(\omega)$ which are bounded in $\omega$ lead to different decompositions of $M(\omega)$, and can be useful as well. In particular, if $M(\omega)$ is a bistochastic matrix, $\psi(\omega)$ can be chosen as an $M^*(\omega)$-invariant vector which is independent of $\omega$.

\subsection{A Law of Large Numbers}

\medskip
\noindent
We now turn to the asymptotics of the Markov process \fer{m16}. 

\begin{thm}[Fluctuating process]
\label{thm2}
Let $M(\omega)$ be a random reduced dynamics operator s.t. that $\p(M(\omega)\in\cme)>0$. There exists a set $\Omega_2\subset\Omega^{\N^*}$ s.t. ${\mathbb P}(\Omega_2)=1$ and, for all $\po\in\Omega_2$,
\begin{equation}
\lim_{N\rightarrow\infty}\frac 1N\sum_{n=1}^N\theta_n(\po) = \theta,
\label{m21}
\end{equation}
where 
\begin{equation}
\theta = \lim_{n\ra\infty}{\mathbb E}[\theta_n]
=P^*_{1,{\mathbb E}[M]}{\mathbb E}[\psi]=P^*_{1,{\mathbb E}[M]}\psi_\cS.
\label{m21.01}
\end{equation}
\end{thm}

\medskip
\noindent {\it Remarks:} 1.
 The ergodic average limit of $\theta_n(\po)$ does not depend on the particular choice of $\psi(\omega)$. This follows from the last equality in \fer{m21.01}.

2. The second  equality in (\ref{m21.01}) stems from
\be
{\mathbb E}[\theta_n]=\sum_{k=0}^{n-1}({\mathbb E}[M_Q])^k{\mathbb E}[\psi],
\ee
by independence, and which converges to $P_{1, {\mathbb E}[M]}^*{\mathbb E}[\psi]$ by (\ref{pem}). The third equality follows from (\ref{idmoy}).

3. Comments on the proof of these Theorems are provided below.

\medskip
\noindent
Combining Theorems \ref{thm1} and \ref{thm2} we immediately get the following result.
\begin{thm}[Ergodic theorem for RRDP]
\label{thm3}
Let $M(\omega)$ be a random reduced dynamics operator. Suppose  $\p(M(\omega)\in\cme)>0$. Then there exists a set $\Omega_3\subset\Omega^{\N^*}$ s.t. ${\mathbb P}(\Omega_3)=1$ and, for all $\po\in\Omega_3$,
\begin{equation}
\lim_{N\rightarrow\infty}\frac 1N\sum_{n=1}^N M(\omega_1)\cdots
M(\omega_n) =|\psi_\cS\ket\bra \theta|=P_{1,{\mathbb E}[M]}.
\label{m21.1}
\end{equation}
\end{thm}

\noindent {\it Remarks\ } 1. If one can choose $\psi(\omega)\equiv\psi$ to be independent of $\omega$, then we have by  (\ref{eq:theta2}) that $\theta_n(\po)=\psi$, for all $n,\po$. Thus,  from \fer{m15}-\fer{m17}, we get the stronger result
$
\lim_{n\rightarrow\infty} M(\omega_1)\cdots
M(\omega_n)= |\psi_\cS\ket\bra \psi|$, a.s., exponentially fast.

2. This result can be viewed as a strong law of large numbers for the matrix valued process $\Psi_n(\po)=M(\omega_1)\cdots M(\omega_n)$.\medskip

\noindent{\bf Comments:}
The existence of (ergodic) limits of products of random operators is known for a long time and under very general conditions, see e.g. \cite{beckschwarz} ,  \cite{k} . However, the explicit value of the limit depends on the detailed properties of the set of random matrices considered. The point of our analysis is thus the explicit determination of the limit   (\ref{m21.1}) which is crucial for the applications to the dynamics of random repeated interaction quantum systems. 

The more difficult part of this task is to prove Theorem \ref{thm1}. The idea consists  in identifying matrices in the product $\Psi_n(\po)$ which are equal (or close) to a fixed matrix $M$ that belongs to $\cme$. Consecutive products of $M$ give an exponential decay, whereas products of other matrices are uniformly bounded. Then one shows that the density of long strings of consecutive $M$'s in a typical sample is finite.
Once this is done, a self-contained proof of Theorem \ref{thm3}, is not very hard to get  \cite{bjm2} .\\

On the other hand, given Theorem \ref{thm1} and the existence result of \cite{beckschwarz} , we can  deduce Theorem \ref{thm3} as follows.  Let us state 
the result of Beck and Schwarz in our setup. Let $T$ denote the usual shift operator on $\Omega^{{\mathbb N}^*}$ defined by $(T\po)_j=\po_{j+1}$, $j=1,2, \cdots$.
\begin{thm}[Beck and Schwartz\cite{beckschwarz}] Let $M(\omega)$ be a random reduced dynamics operator on $\Omega$.  Then there exists a matrix valued random variable  
$L(\po)$ on $\Omega^{{\mathbb N}^*}$, s.t.   ${\mathbb E}[\|L\|]<\infty$, which satisfies almost surely
\be\label{proT}
L(\po)=M(\omega_1)L(T\po),
\ee
where $T$ is the shift operator, and 
\be
\lim_{N\rightarrow\infty}\frac 1N\sum_{n=1}^N M(\omega_1)\cdots
M(\omega_n) =L(\po).
\ee
\end{thm}
Further assuming  the hypotheses of Theorem \ref{thm1},  and making use of 
the decomposition (\ref{m15}), we  get that $L$ can be written as 
\be
L(\po)=|\Psi_{\cS}\ket\bra \theta(\po)|,
\ee
for some random vector $\theta(\po)$.  Now, due to (\ref{proT}) and the fact that $\psi_\cS$ is invariant, $\theta(\po)$ satisfies 
\be
\theta(\po)=\theta(T\po) \ \ \mbox{a.s.}
\ee
The shift being ergodic, we deduce that $\theta$ is constant a.s., so that
\be
\theta(\po)={\mathbb E}[\theta] \ \ \mbox{a.s.}
\ee
which, in turn, thanks to Proposition \ref{prop:unifbound} and Lebesgue dominated convergence Theorem, allows to get from (\ref{m16}) 
\be
{\mathbb E}[\theta]=\lim_{n\ra\infty}{\mathbb E}[\theta_n]=P^*_{1,{\mathbb E}[M]}\Psi_{\cS}.
\ee

\subsection{Limit in Law and Lyapunov Exponents}\label{ssec:met}

We present here results for products ``in reverse order'' of the form 
$\Phi_n(\po):= M(\omega_n) \cdots M(\omega_1)$, which have the same law as $\Psi_n(\po)$.
They also yield information about the Lyapunov exponent of the process. The following results are standard, see  e.g. \cite{la} . The limits 
\[
\Lambda_\Phi(\po)=\lim_{n\ra\infty}(\Phi_n(\po)^*\Phi_n(\po))^{1/2n} \ \  {\rm\  and\ } \ \ 
\Lambda_\Psi(\po)=\lim_{n\ra\infty}(\Psi_n(\po)^*\Psi_n(\po))^{1/2n}
\]
exist almost surely, the top Lyapunov exponent $\gamma_1(\po)$ of $\Lambda_\Phi(\po)$ 
coincides with that of $\Lambda_\Psi(\po)$, it  is constant a.s., and so is its multiplicity. It is in general difficult to prove
that the multiplicity of $\gamma_1(\po)$ is $1$. 
\begin{thm}
\label{prop:randlyap} 
Suppose $\p(M(\omega)\in\cme)>0$. Then there exist $\alpha>0$, a random vector 
\be
\eta_\infty(\po)=\lim_{n\ra\infty}\psi(\omega_1)+
M_{Q}^*(\omega_1)\psi(\omega_{2})+\cdots +M_{Q}^*(\omega_1)\cdots M_{Q}^*(\omega_{n-1}) \psi(\omega_n)
\ee 
and $\Omega_4\subset\Omega^{{\mathbb N}^*}$ with $\P(\Omega_4)=1$ such that for any $\po\in\Omega_4$ and $n\in\N^*$
\be\label{phicvg}
\Big\| \Phi_n(\po)- |\psi_\cS\ket\bra\eta_\infty(\po)| \Big\| \leq C_{\po} e^{-\alpha n}, \ \ \mbox{for some } \  C_{\po}.
\ee
As a consequence, for any  $\po\in\Omega_4$, $\gamma_1(\po)$ is of multiplicity one. 
\end{thm}

\noindent{\bf Comments:}
While the Theorems above on the convergence of asymptotic states give us 
the comfortable feeling provided by almost sure results, it is an important 
aspect of the theory to understand the fluctuations around the asymptotic 
state the system reaches almost surely. In our iid setup, the law of the 
product $\Psi_n(\po)$ of RRDO's coincides with the law of $\Phi_n(\po)$ which
converges exponentially fast to $ |\psi_\cS\ket\bra\eta_\infty(\po)|$. 
Therefore, the fluctuations are encoded in the law of the random vector 
$\eta_\infty(\po)$.  It turns out it is quite difficult, in general, to get informations about this law. There are partial results only about certain aspects of the law of such random vectors in case they are obtained by means of matrices belonging to some subgroups of $Gl_d(\mathbb R)$ satisfying certain irreducibility conditions, see e.g. \cite{dsetal}. However, these results do not apply to our RRDO's.

\subsection{Generalization}

A generalization of the analysis performed for observables acting on $\cS$ only described above allows to establish the following corresponding results when instantaneous observables are considered.

The asymptotics of the dynamics \fer{intro1'}, in the random case, is encoded in the product 
\[
M(\omega_1)\cdots M(\omega_{m-l-1})N(\omega_{m-l},\ldots,\omega_{m+r}),
\]
where $N:\Omega^{r+l+1}\ra M_d({\mathbb C})$ is given in assumption (R2).

\begin{thm}[Ergodic limit of infinite operator product] \ \\
\label{thmintro1} Assume $M(\omega)$ is a RRDO and (R2) is satisfied. Suppose that $\p(M(\omega)\in\cme)\neq 0$. Then ${\mathbb E}[M]\in\cme$. Moreover, there exists a set ${\Omega}_5 \subset
 \Omega^{\N^*}$ of probability one 
s.t. for any $\po=(\omega_n)_{n\in\mathbb N}\in{\Omega}_5$,
\begin{equation*}
\lim_{\nu\rightarrow\infty}\frac 1\nu\sum_{n=1}^\nu M(\omega_1)\cdots
M(\omega_n) N(\omega_{n+1},\ldots,\omega_{n+l+r+1})= |\psi_\cS\ket\bra \theta|\ \E[N],
\end{equation*}
where $\theta = P^*_{1,{\mathbb E}[M]}\psi_\cS$.
\end{thm}

As in the previous Section, a density argument based on the cyclicity and separability of the 
reference vector $\psi_0$ allows to obtain from Theorem \ref{thmintro1} the asymptotic state 
for all normal initial states $\varrho$ on $\fm$ given as  Theorem \ref{thmintro2}


\bigskip
{\bf Acknowledgements.\ }

I wish to thank I. Beltita, G. Nenciu and R. Purice who organized the 10th edition of  QMath for their kind invitation and  L. Bruneau and M. Merkli for a very enjoyable collaboration.


\end{document}